\documentclass[a4paper]{jpconf}
\usepackage{graphicx}
\usepackage{amsfonts}
\usepackage{amssymb}
\usepackage{latexsym}
\usepackage{cite}
\newcommand{\be}{\begin{equation}}
\newcommand{\ee}{\end{equation}}
\newcommand{\bea}{\begin{eqnarray}}
\newcommand{\eea}{\end{eqnarray}}
\newcommand{\nn}{\nonumber}
\newcommand{\ba}[1]{\begin{array}{#1} }
\newcommand{\ea}{\end{array}} 
\newcommand{\expval}[1]{\mbox{$\langle #1 \rangle$}}
\newcommand{\del}{\partial}
\newcommand{\half}{\frac{1}{2}}
\newcommand{\halfsm}{\mbox{$\scriptstyle \frac{1}{2}$}}
\newcommand{\sub}[1]{_{\scriptscriptstyle #1}}
\newcommand{\sups}[1]{^{\scriptscriptstyle #1}}
\newcommand{\colvec}[2]{\mbox{$\left( \ba{c}\displaystyle #1 \\
\displaystyle #2 \ea \right)$ }}
\newcommand{\ket}[1]{|#1\rangle}
\newcommand{\bra}[1]{\langle #1|}
\newcommand{\lket}[1]{\bigl|#1\bigr\rangle}
\newcommand{\lbra}[1]{\bigl\langle #1\bigr|}

\newcommand{\delslash}{\partial\!\!\!/}
\newcommand{\eqn}[1]{Eq.~(\ref{eq:#1})}
\newcommand{\fig}[1]{Fig.~\ref{fig:#1}}

\newcommand{\mpi}{m\sub{\pi}}
\newcommand{\LD}{{\cal L}}

\newcommand{\psibar}{\bar{\psi}}

\newcommand{\scprod}[2]{\mbox{$\vec{#1}\cdot\vec{#2}$}}

\newcommand{\vpi}{\mbox{$\vec{\pi}$}}
\newcommand{\fpi}{f\sub{\pi}}

\newcommand{\aplus}{\alpha\sub{+}}
\newcommand{\aminus}{\alpha\sub{-}}

\newcommand{\gA}{g\sub{A}}

\newcommand{\chih}{\chi\sub{h}}
\newcommand{\ap}{\alpha\sub{+}}
\newcommand{\am}{\alpha\sub{-}}
\newcommand{\cR}{\cosh R\mpi}
\newcommand{\sR}{\sinh R\mpi}
\newcommand{\crm}{\cosh r\mpi}
\newcommand{\srm}{\sinh r\mpi}
\newcommand{\cL}{\cosh L\mpi}
\newcommand{\sL}{\sinh L\mpi}
\newcommand{\gAhh}{g\sub{A}\sups{\textrm{hh}}}

\newcommand{\mev}{\,{\rm MeV}}
\newcommand{\fm}{\,{\rm fm}}

\begin{document}

\title{Finite Volume Dependence of Hadron Properties and Lattice QCD}

\author{Anthony W.~Thomas$^1$, Jonathan D.~Ashley$^2$, 
Derek B.~Leinweber$^2$
and Ross D.~Young$^1$}

\address{$^1$Jefferson Lab, 12000 Jefferson Ave., Newport News VA 23606 USA}
\address{$^2$CSSM, Department of Physics, 
University of Adelaide, Adelaide SA 5005 Australia}


\begin{abstract}
Because the time needed for a simulation in lattice QCD varies 
at a rate exceeding the
fourth power of the lattice size, it is important to understand how
small one can make a lattice without altering the physics beyond
recognition. It is common to use a rule of thumb that the pion mass
times the lattice size should be greater than (ideally much greater
than) four (i.e., $m_\pi L \gg 4$).
By considering a relatively simple chiral quark model we are led to
suggest that a more realistic constraint would be $m_\pi (L - 2R) \gg 4$,
where $R$ is the radius of the confinement region, which
for these purposes could be taken to be around 0.8-1.0 fm. Within the
model we demonstrate that violating the second condition can lead to
unphysical behaviour of hadronic properties as a function of pion mass.
In particular, the axial charge of the nucleon is found to decrease
quite rapidly as the chiral limit is approached.
\end{abstract}

\section{Introduction}
Our current capacity to compute hadron properties in lattice QCD is
constrained by the need to take many limits. For example, we need to
take the spacing $a \rightarrow 0$, the size of the lattice $L
\rightarrow \infty$ and the quark mass to around 5 MeV. Of course, these
limits are not unconnected because the size of a region of space big
enough to contain (say) a proton will need to grow with the Compton
wavelength of the pion. At present, with lattice spacings that represent
a reasonable approximation to the continuum limit and for full QCD with
reasonable chiral symmetry, the time for a lattice simulation scales roughly
like $m_\pi^{-9}$ and we are limited to pion masses larger than 0.4-0.5
GeV. This has led to much effort to explore the application of chiral
perturbation theory as a tool for extrapolating hadron properties to the
physical pion mass~\cite{Young:2002ib,Leinweber:2003dg,Procura:2003ig}.

In such an environment there is great interest in seeing whether one can
lower the pion mass without increasing the size of the lattice, thereby
saving a factor of $m_\pi^4$. Indeed, there have been many calculations
for which even the rather optimistic rule of thumb that $m_\pi L > 4$
has not been satisfied. Considerable attention has been devoted to
studying such systems within the framework of effective field theory in
order to understand how the relevant path integral might change as we go
from one regime to another~\cite{Detmold:2004ap,Beane:2004rf}.

The question we ask is somewhat different. We consider a simple chiral quark
model upon which we impose boundary conditions which roughly approximate
those on a lattice. Simple inspection of the solutions naturally leads
one to conclude that the condition noted above is incorrect. Indeed, the
pion cloud of the nucleon does not even begin until one is outside the
region of space in which the valence quarks are confined. Within the bag
model this is characterised by the bag radius, $R$, and within the
cloudy bag model as well as the model considered here, this radius is
where the pion field peaks. The asymptotic behaviour of the pion field
is therefore $\phi(r) \sim exp[- m_\pi (r - R)] /r^2$ and the correct
condition for the pion field to be small on a spherical boundary
surrounding the nucleon is that $(L/2 -R) \gg m_\pi^{-1}$, with $L$ the
diameter of the spherical ''lattice''. (Note that in sects. 2 and 3 we
will use $L$ to denote the {\it radius} of such a region.)

If we consider a typical case where the pion mass is large, say $m_\pi =
600$ MeV, then on a 2 fm lattice $m_\pi L \sim 6$ and the commonly
quoted condition would suggest that we had a sufficiently large lattice.
However, with a bag radius of order 0.8 fm, we find $m_\pi (L/2 -R) \sim
0.6$ and the pion field has no chance to drop to zero before we reach
the edge of the lattice! Indeed, even at this relatively large mass a
3 fm lattice would be a minimal requirement for full QCD simulations. 
{}From the mathematical point
of view, having the boundary too close means that we are not restricted
to the well behaved solution of the second order differential equation
but can have a significant coefficient for the divergent solution.
Through the coupling to the confined valence quarks this can in turn
change the internal valence structure of the hadron.

Considerations such as these
help us to understand why some hadronic properties  
exhibit a dramatic volume dependence as the pion mass is varied.
Perhaps the most famous example is the axial charge of the nucleon
where early simulations revealed a striking decrease of $g_A$ as the
pion mass decreased --- in the opposite direction from the experimental
data~\cite{Sasaki:2001th,Jaffe:2001eb,Cohen:2001bg}. 
We shall see that our simple chiral model is able to reproduce
this feature.

\section{A Simple Chiral Quark Model}
One of the earliest attempts to restore chiral symmetry to the MIT bag 
model was made in 1975 by Chodos and Thorn~\cite{Chodos:1975ix}. 
Their solution, 
a precursor to the CBM~\cite{Miller:1979kg,Thomas:1982kv}
was a straight forward generalization of the linear 
sigma model, with pion and sigma fields coupling linearly to the confined 
quarks at the bag boundary. As well as presenting a perturbative solution 
in the pion field (as in the CBM), Chodos and Thorn attempted to find 
an exact solution to the resulting equations of motion. The only case where 
this was feasible was for a highly idealized baryon called the 
``hedgehog'', in which the 3 confined quarks all inhabit the same mixed 
spin-flavour singlet, and which hence is not an eigenstate of either 
isospin or angular momentum. As a result, the hedgehog solution does not 
correspond to any physical particle. 
Despite this fact, the hedgehog approximation does provide a simple 
classical solution with which we can then study a number of 
phenomena with relative ease. 

Chodos and Thorn suggest the following chirally invariant Lagrangian density 
based on the MIT bag model Lagrangian: 
\be\label{eq:hh L CT}
    \LD\sub{\mathit{CT}} =
    [\psibar i\delslash\psi - B]\theta_V - 
    \lambda \psibar(\sigma+i\scprod{\tau}{\pi}\gamma_5)\psi\,\delta_S 
     +\half(\del_\mu\sigma)(\del^\mu\sigma) + \half
    (\del_\mu\vpi)\cdot(\del^\mu\vpi),
\ee
where $\psi$, $\vec{\pi}$ and $\sigma$ are the quark, pion and sigma fields, 
respectively, and $\lambda$ is a Lagrange multiplier which turns out to be 
$\halfsm(\sigma^2 + \pi^2)^{-1/2}$. 
This is invariant under the appropriate infinitesimal chiral transformation 
and the corresponding axial current:
\be\label{eq:hh ax}
    \vec{A}^\mu = 
    \half \psibar\vec{\tau}\gamma_5\gamma^\mu\psi\theta_V 
    + (\del^\mu\sigma)\vpi - \sigma(\del^\mu\vpi), 
\ee
is conserved. 

To this Lagrangian we will also add explicit chiral symmetry breaking 
quark and pion mass terms. 
To generate a pion mass we could follow the method of the linear sigma model, 
spontaneously breaking chiral symmetry and then tipping the Mexican hat 
potential, generating masses for both the $\vpi$ and $\sigma$ fields. 
However, this process leads to some very complex equations of motion for the 
pion and sigma fields, destroying the simplicity of Chodos and Thorn's 
classical solution. 
{}For simplicity we include the pion mass (and the corresponding quark
mass) by hand, leaving the 
$\sigma$ field massless. 
The resulting Lagrangian is
\bea
    \LD &=& 
    [\psibar(i\delslash-m_q)\psi - B]\theta_V - 
    \lambda \psibar(\sigma+i\scprod{\tau}{\pi}\gamma_5)\psi\,\delta_S \nn\\
    && +\half(\del_\mu\sigma)(\del^\mu\sigma) + \half
    (\del_\mu\vpi)\cdot(\del^\mu\vpi) - \half\mpi^2\vpi\cdot\vpi.
\label{eq:hh L covariant}
\eea

Minimizing the action leads to five Euler-Lagrange equations. 
For a static spherical bag of radius $R$ and static $\sigma$ and $\vpi$ 
fields they are as follows: 
\bea
    (i\delslash-m_q)\psi = 0,&& r\leq R;\label{eq:hheq dirac}\\
    i\hat{r}\cdot\vec{\gamma}\psi = -\xi
    (\sigma+i\vec{\tau}\cdot\vpi\gamma_5)\psi,&& r=R;\label{eq:hheq lbc}\\
    \nabla^2\sigma = \half\xi 
    \psibar\psi\,\delta(r-R);&&\label{eq:hheq sig}\\
    \nabla^2\vpi - \mpi^2\vpi = \half\xi 
    \psibar i\vec{\tau}\gamma_5\psi\,\delta(r-R);&&\label{eq:hheq pi}\\
    B = -\half\xi \frac{\del}{\del r} \bigl[
    \psibar(\sigma+i\scprod{\tau}{\pi}\gamma_5)\psi\bigr]_{r=R,}&&
    \label{eq:hheq nlbc}
\eea
where we use the notation $\xi=[\sigma^2(R) + \pi^2(R)]^{-1/2}$. 
The first of these equations is just the Dirac equation for the confined 
quarks, whilst 
Eqs.~(\ref{eq:hheq sig}) and (\ref{eq:hheq pi}) are equations of motion for 
the sigma and pion fields. There are also two boundary conditions --- the linear 
boundary condition, \eqn{hheq lbc}, and the non-linear boundary condition, 
\eqn{hheq nlbc}, with similar roles to their roles in the MIT model. 

Chodos and Thorn solve this system of equations (without quark and pion 
masses) on an infinite volume. 
Following their method, we will solve the system on a finite spherical volume 
of radius $L$. 
Ideally we would use a rectangular volume similar to that used in lattice 
simulations but because of the radial nature of the solutions it is 
natural to use a spherical volume and any attempt to solve on a rectangular 
volume would be significantly more complicated. 

The quark field solution is just the MIT quark wavefunction, 
\be\label{eq:hh q soln}
    \psi(\vec{r},t) = \colvec{\aplus j_0\bigl(\frac{\Omega r}{R}\bigr)}
    {i\aminus j_1\bigl(\frac{\Omega r}{R}\bigr)\,\vec{\sigma}\cdot\hat{r} }
    \chih \,\theta(R-r) \,e^{-i\alpha t/R},
\ee
where we define $\alpha=ER$ and $\Omega=kR$ (where $k$ is to be determined by
the appropriate eigenvalue condition) related 
by the equation 
\be
    \alpha = \sqrt{\Omega^2 + (m_qR)^2},
\ee
and 
\be
    \alpha\sub{\pm} = \sqrt{\frac{\alpha \pm m_qR}{\alpha}}. 
\ee

The defining property of the hedgehog solution is the choice of static, 
radially dependent sigma and pion fields, 
\bea
    \vpi(\vec{r}) &=& g(r)\hat{r},\label{eq:hh pi form}\\
    \sigma(\vec{r}) &=& f(r)\label{eq:hh sig form},
\eea
and the choice of spinor-isospinor, $\chih$, proportional to a mixed 
spin-isospin singlet state,
\be
    \hat{\chih} = 
    \frac{1}{\sqrt{2}}\bigl(\ket{d\uparrow}-\ket{u\downarrow}\bigr), 
\ee
so that it has the property, 
\be
    (\vec{\sigma}+\vec{\tau})\chih = 0.
\ee
These choices greatly simplify the equation for the pion field by 
ensuring that $\psibar\vec{\tau}\gamma_5\psi$ is proportional to $\hat{r}$ 
and hence to $\vpi$:
\be
    \psibar\vec{\tau}\gamma_5\psi = -2i\,j_0 j_1\,\chih^\dagger\chih\,\hat{r}.
\ee
Substituting Eqs.~(\ref{eq:hh q soln}), (\ref{eq:hh pi form}) and 
(\ref{eq:hh sig form}) into the equations of motion for $\vpi$ and 
$\sigma$, yields the following two second order differential equations 
in $f$ and $g$: 
\bea\label{eq:f de}
    f''(r) + \frac{2}{r}f'(r) &=& \half \xi \bigl(\ap^2j_0^2(\Omega)
    -\am^2j_1^2(\Omega)\bigr)\chih^\dagger\chih\,\delta(R-r) 
    \nn\\&\equiv& a\,\delta(R-r),\\
    g''(r) + \frac{2}{r}g'(r) -\bigl(\frac{2}{r^2}+\mpi^2\bigr)g(r) &=& 
    \xi\ap\am j_0(\Omega)j_1(\Omega)\chih^\dagger\chih \,\delta(R-r)
    \nn\\ &\equiv& b\,\delta(R-r).
    \label{eq:g de}
\eea
Enforcing regularity at the origin, Eqs.~(\ref{eq:f de}) and 
(\ref{eq:g de}) have solutions 
\bea
    f(r) &=& f_0 + aR^2\bigl(\frac{1}{R}-\frac{1}{r}\bigr)\theta(r-R),\\
    g(r) &=& 
    \bigl(-\frac{\crm}{r\mpi}+\frac{\srm}{r^2\mpi^2}\bigr) 
    \Bigl[C + b\bigl(R\sR - \frac{\cR}{\mpi}\bigr)\theta(r-R)\Bigr] + \nn\\&&
    \bigl(\frac{\srm}{r\mpi}-\frac{\crm}{r^2\mpi^2}\bigr)
    b\bigl(R\cR-\frac{\sR}{\mpi}\bigr)\theta(r-R),
\eea
where $f_0$ and $C$ are constants of integration. 

At the boundary $r=L$ 
we set the derivative $\vec{\pi}'$ to zero to resemble the 
periodic boundary conditions applied in lattice simulations. {\it We stress
that in this context $L$ is not the lattice length, which would be
of order $2L$.} Because there 
is no sigma mass we cannot enforce a similar condition on $\sigma$. Instead 
we leave $f_0$ as a free parameter for the moment. 
The resulting solutions are 
\bea
    f(r) &=& f_0 + aR^2(\frac{1}{R}-\frac{1}{r})\theta(r-R),\\
    g(r) &=& b\bigl(R\cR-\frac{\sR}{\mpi}\bigr)
    \bigl(\frac{\srm}{r\mpi}-\frac{\crm}{r^2\mpi^2}\bigr)\theta(r-R) + \nn\\&&
    b\bigl(R\sR - \frac{\cR}{\mpi}\bigr)
    \bigl(\frac{\crm}{r\mpi}-\frac{\srm}{r^2\mpi^2}\bigr)\theta(R-r) + \nn\\&&
    b\bigl(R\cR-\frac{\sR}{\mpi}\bigr) 
    \bigl(-\frac{\crm}{r\mpi}+\frac{\srm}{r^2\mpi^2}\bigr) \Gamma(L\mpi),
\eea
where the $L$-dependence is given by the function, 
\be
    \Gamma(L\mpi) = \biggl(
    \frac{\frac{\cL}{L\mpi}-2\frac{\sL}{L^2\mpi^2}+2\frac{\cL}{L^3\mpi^3}}
    {\frac{\sL}{L\mpi}-2\frac{\cL}{L^2\mpi^2}+2\frac{\sL}{L^3\mpi^3}}
    \biggr). 
\ee
(Note that $\Gamma(L\mpi)\to1$ as $L\to\infty$). 

\subsection{The linear boundary condition}
The solutions to Eqs.~(\ref{eq:hheq dirac}), (\ref{eq:hheq sig}) and 
(\ref{eq:hheq pi}) are dependent on the parameters $\Omega$, $R$, $L$, $m_q$ 
and $\mpi$ but also on the unknown values $f(R)$ and $g(R)\equiv g_0$, which 
are given by 
\bea
    f(R) &=& f_0, \\
    g(R) &=& -\xi \ap\am j_0 j_1 \chih^\dagger\chih 
    \bigl(R\cR-\frac{\sR}{\mpi}\bigr) \times \nn\\&&
    \Bigl[\bigl(-\frac{\sR}{R\mpi}+\frac{\cR}{R^2\mpi^2}\bigr) + 
    \bigl(\frac{\cR}{R\mpi}-\frac{\sR}{R^2\mpi^2}\bigr)\Gamma(L\mpi)\Bigr]\\
    &\equiv& -\xi \, d(\Omega,m_q,\mpi,R,L).
    \label{eq:gxid}
\eea
If we now apply the linear boundary condition, \eqn{hheq lbc}, 
we may express $f_0^2$, 
$g_0^2$ and $1/\xi^2$ in terms of $y$ and $d$: 
\be
    f_0^2 = \frac{4y^2}{1-y^4}\,d, \hspace{1cm}
    g_0^2 = \frac{1-y^2}{1+y^2}\,d, \hspace{1cm}
    \frac{1}{\xi^2} = \frac{1+y^2}{1-y^2}\,d,
\ee
so that our solution now depends only on the parameters $\Omega$, $m_q$, 
$\mpi$, $R$ and $L$.

\subsection{The non-linear boundary condition}
The last condition to apply is the non-linear boundary condition, 
\eqn{hheq nlbc}, which sets the bag energy density $B$. Because of the 
discontinuity in the derivatives of the 
$\vpi$ and $\sigma$ fields at the bag boundary, $r=R$, we take the average 
of the derivative on each side\footnote{With this prescription, Chodos and 
Thorn found that the non-linear boundary condition (n.l.b.c.\  ) 
corresponded with conservation of energy in 
the hedgehog.}, that is:  
\be
    \frac{\del \sigma}{\del r}\Bigr|_{r=R} \equiv 
    \frac{1}{2} \Bigl(\frac{\del \sigma^{(in)}}{\del r} + 
    \frac{\del \sigma^{(out)}}{\del r}\Bigr)\Bigr|_{r=R}.
\ee
With the solutions for $\psi$, $\vpi$ and $\sigma$, \eqn{hheq nlbc} leads 
to an expression for $B$ of the form, 
\be\label{eq:hh nlbc simple}
    4\pi R^4 B = \tau(\Omega,m_q,\mpi,R,L), 
\ee
where $\tau$ is a complicated function of the bag frequency $\Omega$, 
hedgehog and volume radii ($R$ and $L$ respectively), and quark and 
pion masses. 
Because of its complexity, the explicit form of $\tau$ and the 
involved steps required to find it are not shown. 

\begin{figure}[ht]
\begin{center}
\includegraphics[width=8cm]{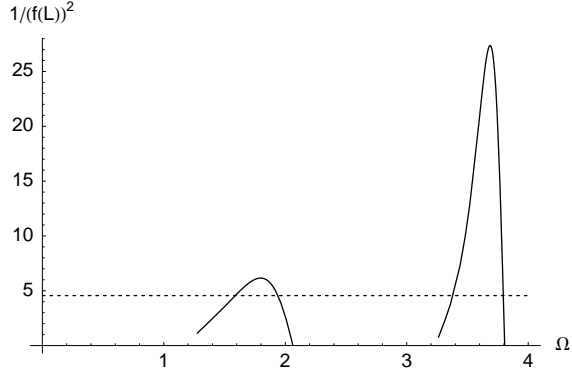}
\caption{Plot of the inverse square of the value of the sigma field at the 
boundary $r=L$ versus the bag frequency $\Omega$. 
The horizontal line represents 
the boundary condition $f(L)=\fpi$.}
\label{fig:fL 100}
\end{center}
\end{figure}

Once the masses and volume size are set, \eqn{hh nlbc simple} becomes a 
three-way relationship between the parameters $B$, $R$ and $\Omega$. 
By choosing a value for the bag energy density (a property of the vacuum) 
the radius $R$, and hence the entire solution, will be completely determined 
by $\Omega$. Using this fact we can now find an eigenvalue condition for 
$\Omega$.
Using the linear and non-linear boundary conditions, we find that 
the value of the sigma 
field at the boundary $r=L$, 
\be
    \sigma(L) = f(L) = f_0 + a R^2 \bigl(\frac{1}{R}-\frac{1}{L}\bigr), 
\ee
is completely specified by the energy parameter $\Omega$. 
\fig{fL 100} shows a plot of $1/f(L)^2$ against $\Omega$ for fixed 
parameters $B$, $m_q$, $\mpi$ and $L$. 
Physical solutions occur for $\Omega$ within specific allowed regions 
or bands. Outside of these regions either $\sigma^2$ is negative and hence 
$\sigma$ is imaginary, or $\tau(\Omega)$ is negative, in which case there 
are no real solutions to the n.l.b.c. 
By selecting a boundary condition for the sigma field 
(represented by a horizontal line in \fig{fL 100}) we can find 
eigenvalues of the bag frequency, $\Omega$. 
From \fig{fL 100} we see that $1/f(L)^2$ has a limited size so there 
is a limit to how small we can make $\sigma(L)$. 

So what boundary condition should we apply to $\sigma$? 
Starting with the PCAC relation 
\be\label{eq:hh PCAC}
    \lbra{0}\del_\mu A_a^\mu(x)\lket{\pi_b(q)} = 
    \fpi \mpi^2 \,\delta_{ab}\, e^{-iq\cdot x} \, ,  
\ee
we substitute the axial current of \eqn{hh ax} into the left-hand side 
and take the limit $r\to\infty$ (where $\psi=0$) then we find 
\bea
    \lbra{0}\del_\mu A_a^\mu(x)\lket{\pi_b(q)} &=& 
    \lbra{0}\bigl\{ (\del^2\sigma)\pi_a - \sigma(\del^2\pi_a) \bigr\}
    \lket{\pi_b(q)} \\
    &=& \mpi^2\expval{\sigma}\delta_{ab}\,e^{-iq\cdot x} \, .
\eea
Comparing with \eqn{hh PCAC}, we see that the vacuum expectation 
value of the sigma field must be $\expval{\sigma} = \fpi$ 
so we expect the sigma field to go to $\fpi$ at infinity. 

To find the hedgehog solution on an infinite volume we would therefore 
choose $f_0$ such that $f(r)\to\fpi$ as $r\to\infty$. For large volumes, 
setting $f(L)=\fpi$ should still be an appropriate boundary condition. But 
for smaller volumes this is too harsh a condition and is found to introduce 
large volume dependence. As an alternative, we fix the 
sigma field at $r=L$ to the value taken by $\sigma(L)$ in the 
infinite volume solution (in which $\sigma\to\fpi$ as $r\to\infty$). This 
results in a more modest volume dependence, which seems appropriate
given that in non-linear chiral models  
the sigma mass is usually set to infinity and 
we expect volume dependence to arise mainly from the effect of the $\vpi$ 
field.
\begin{figure}[ht]
\begin{center}
\includegraphics[width=10cm]{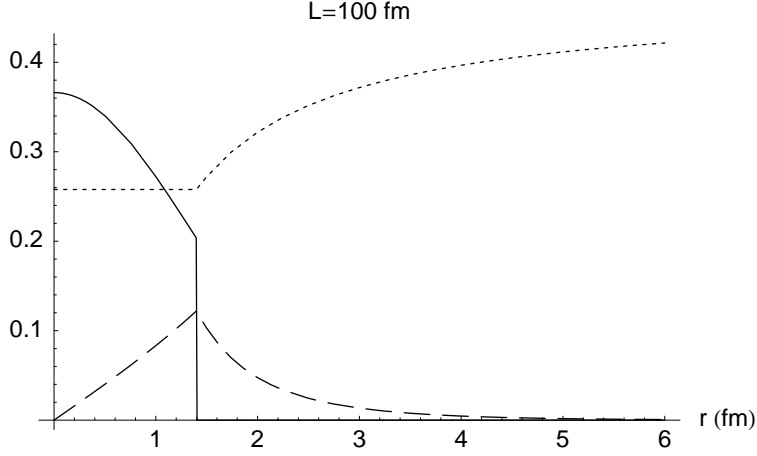}
\caption{The quark density $\psi^\dagger\psi$ (full line) and magnitudes of 
the pion (dashed line) and sigma (dotted line) fields in the hedgehog solution 
calculated at the physical pion mass on a spherical volume of radius 
$L=100$\,fm.}
\label{fig:hh soln 100b}
\end{center}
\end{figure}

The solution displayed in \fig{hh soln 100b} for the quark ground state 
and associated $\vpi$ and $\sigma$ fields was found using the latter 
boundary 
condition on a volume of radius $L=100$\,fm. The quark and pion masses 
were set to $m_q=5$\,MeV and $\mpi=140$\,MeV and the energy density 
$B=13.27$\,MeV$/$fm$^3$ yields a bag radius of $R=1.4$\,fm as 
$L$ goes to infinity. In this case the lowest energy state is given by the 
eigenvalue $\Omega=1.588$. 
Although the choice of bag radius $R=1.4$\,fm may appear slightly large, this 
is in fact the smallest value for which ground state solutions can be found 
on an infinite volume in this model. Clearly what matters in terms of
physics consequences is the difference between $L$ and $R$, rather than
the absolute values. 

\subsection{The mass of the hedgehog}
The energy of the combined system is found by integrating the 
energy-momentum tensor and can be divided into three separate pieces, 
\be
    E\sub{hh} = E\sub{\psi} + E\sub{\mathrm{bag}} + E\sub{\sigma\pi}.
\ee
The quark and bag energies we already know from the MIT model: 
\bea
    E\sub{\psi} &=& \frac{3\alpha}{R},\\
    E\sub{\mathrm{bag}} &=& \frac{4\pi}{3}R^3 B.
\eea
The pion and sigma energy is given by the integral 
\bea
    E\sub{\sigma\pi} &=& \half \int d^3\!r [(\nabla\sigma)^2 
    + (\nabla\vec{\pi})^2 + \mpi^2\vpi^{\,2}] \\
    &=& 2\pi \int_0^L \! dr [r^2 f'(r)^2 + r^2 g'(r)^2 + 
    (2+r^2\mpi^2)g(r)^2].
\eea
{}For the solution presented in \fig{hh soln 100b} the total energy (or 
hedgehog mass) is $998$\,MeV, which is comparable to the physical nucleon mass 
$M\sub{N}\simeq 940$\,MeV.

The axial coupling constant is found by integrating this nucleon matrix 
element of the axial current over all space. In the hedgehog we integrate 
$A\sub{hh}^\mu(x)$ (given by \eqn{hh ax}): 
\bea
    \gAhh\,\delta_{ij} &=& \int d^3\!r \bra{hh}A^i_j(\vec{r})\ket{hh} \\
    &=& \int d^3\!r \bigl[ \half \psibar\tau_j\gamma_5\gamma^i\psi + 
    (\del^i\sigma)\pi_j - \sigma(\del^i\pi_j) \bigr].
\eea
Applying the hedgehog solutions, 
one can obtain closed forms for the quark and meson contributions.

\section{Results: Volume Dependence of the Hedgehog}
\begin{figure}[ht]
\begin{center}
\includegraphics[width=10cm]{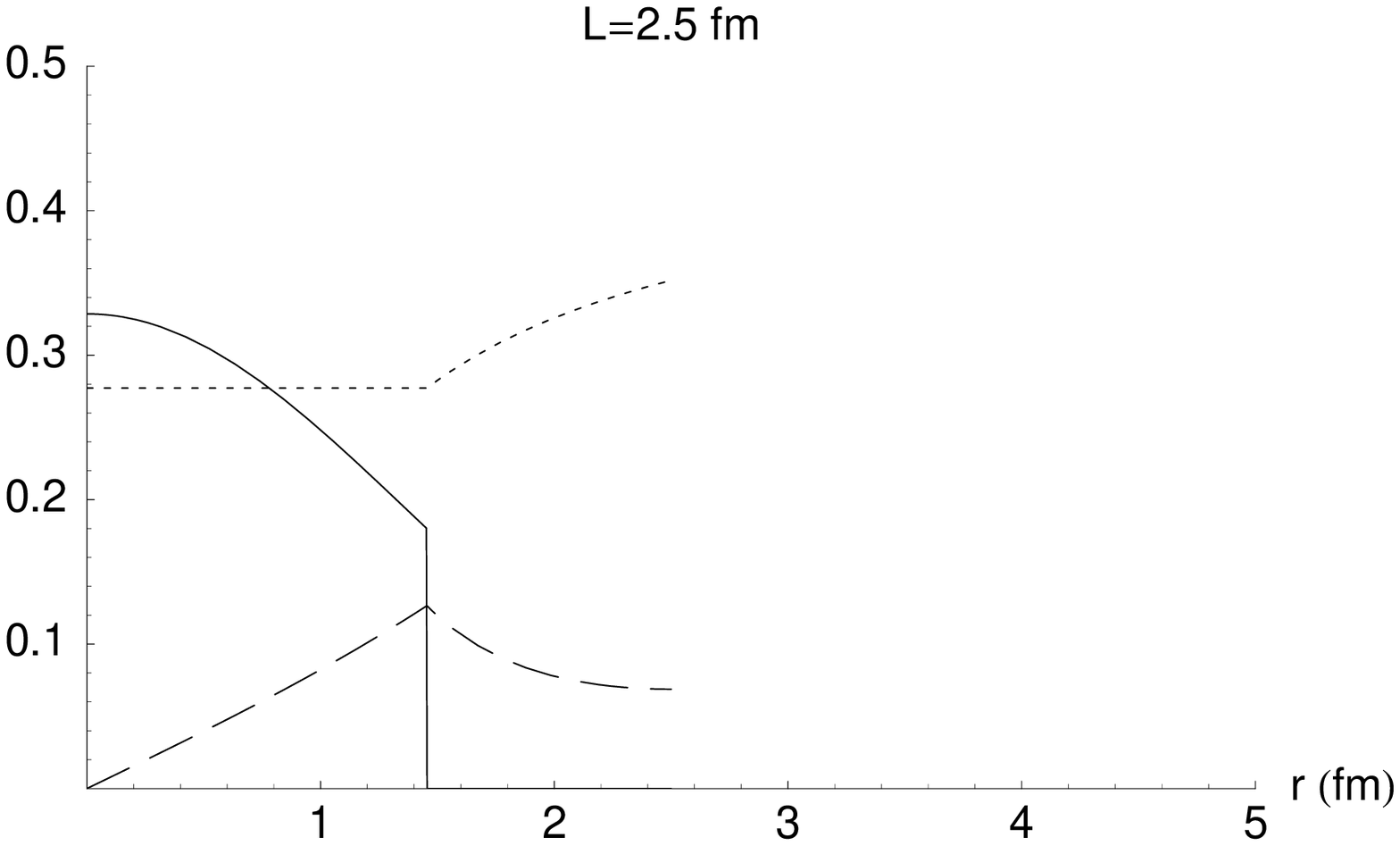} \\
\includegraphics[width=10cm]{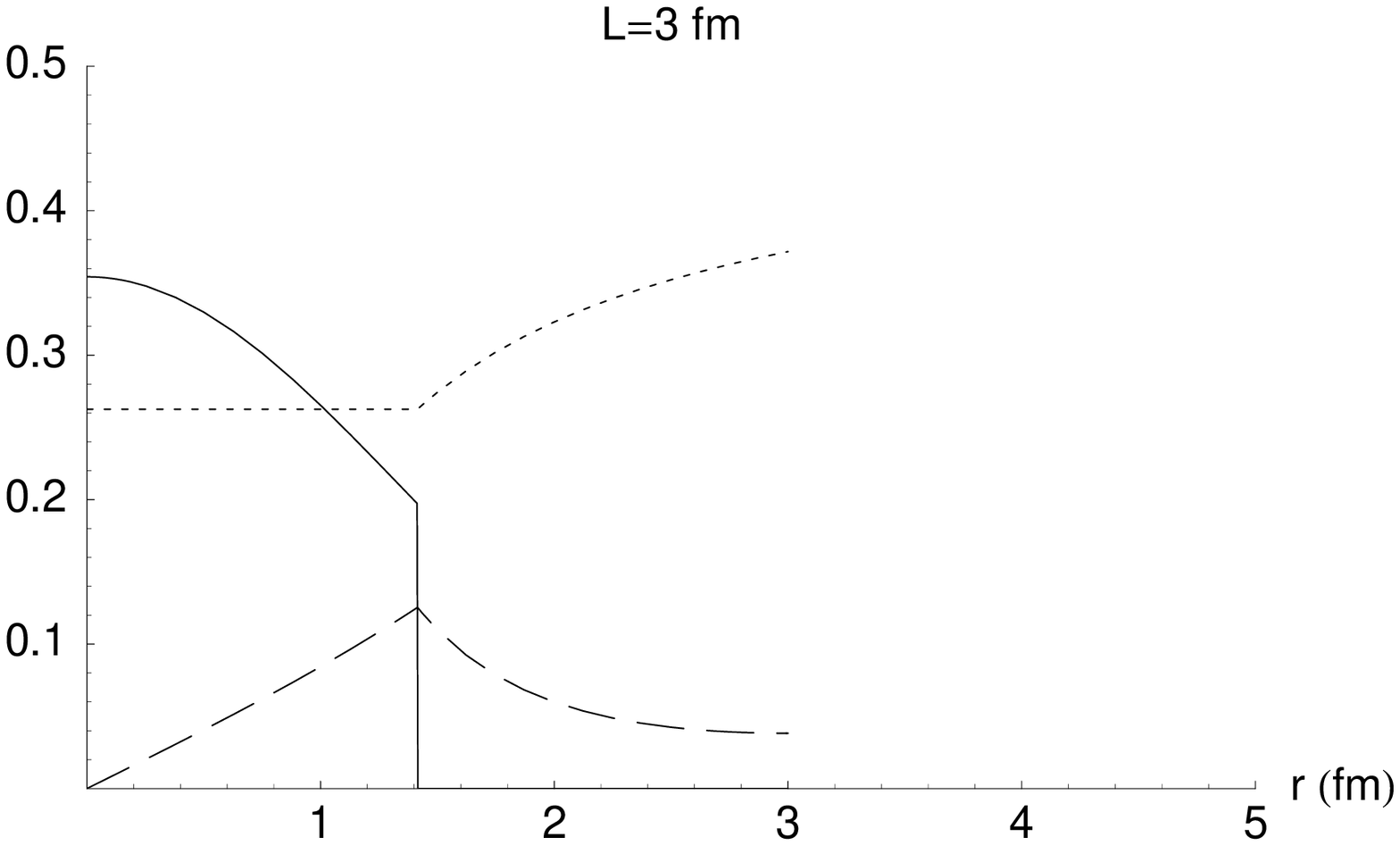}
\caption{The quark density $\psi^\dagger\psi$ (full line) and magnitudes
of
the pion (dashed line) and sigma (dotted line) fields in the hedgehog
solution
calculated at the physical pion mass on spherical volumes of radius
$2.5$ and $3\fm$.}
\label{fig:hh solns}
\end{center}
\end{figure}
Using the technique described in the previous section, we can find solutions 
for the groundstate hedgehog on a range of different sized spherical volumes 
and at different pion and quark masses. For each solution we calculate the 
hedgehog energy and the axial coupling constant. 
The bag energy density is fixed at $B=13.27$\,MeVfm$^{-3}$ 
(to give $R=1.4$\,fm in the infinite volume solution) and the quark and 
pion masses fixed in the proportionality defined by the 
Gell-Mann-Oakes-Renner relation so that 
$m_q = (\mpi/140\mev)^2\;5\mev$.
By choosing the volume size, $L$, and pion mass, $\mpi$, and applying the 
boundary condition on $\sigma$ we find the lowest eigenvalue of $\Omega$, 
which then completely specifies the solution. 

At small pion masses, if the volume becomes too small then there is no 
groundstate solution which satisfies the boundary condition. 
As $L$ decreases, the sigma field will grow whilst the value of $\sigma(L)$ 
required by the boundary condition gets smaller, such that eventually 
there is no value of $\Omega$ for which $\sigma(L)$ is small 
enough.
At the physical pion mass we cannot find solutions on volumes much 
smaller than $L=2.5$\,fm.
For larger pion masses the pion and sigma fields are smaller and we do 
not encounter this problem.

Graphs of the solutions for $2.5$ and $3\fm$ at the physical 
pion mass are shown in Fig.~\ref{fig:hh solns}.
We observe that the hedgehog solutions on different sized 
volumes are relatively similar. Interestingly, as the volume size decreases 
there is a small \emph{increase} in the bag size, accompanied by a 
decrease in the quark density over the bag. There is very little change 
in the $\vpi$ and $\sigma$ fields inside the bag volume. 
{\it The most noticeable change is in the behaviour 
of the pion field outside the 
baryon, which becomes much larger for smaller volumes because there is less 
distance over which it can flatten out to satisfy the periodic boundary 
condition.} 

\begin{figure}[ht]
\begin{center}
\includegraphics[width=13cm]{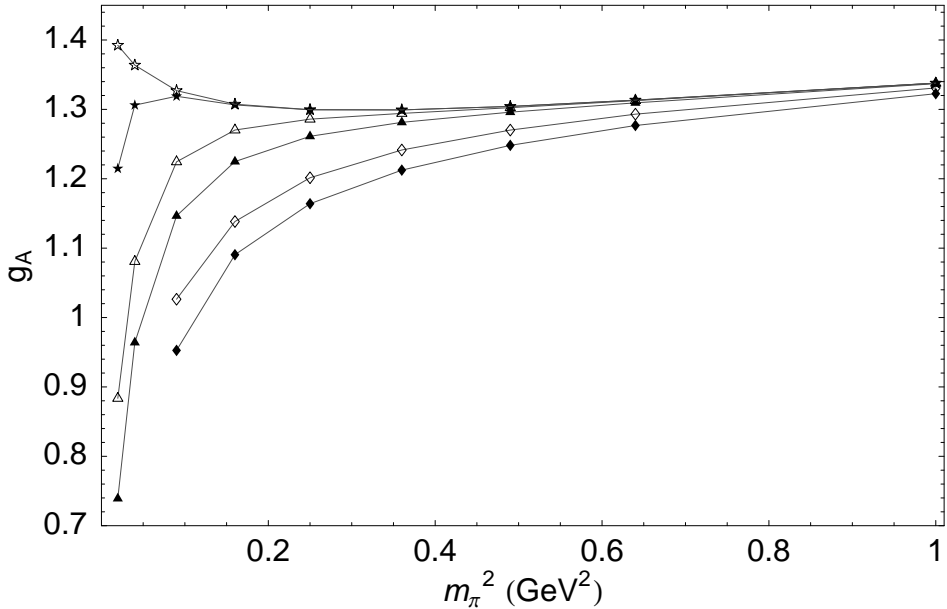}
\caption{Plot of the axial coupling constant $\gA$ in the hedgehog versus 
$\mpi^2$, as calculated on spherical volumes of radius $10$\,fm (open star), 
$5$\,fm (filled star), $3$\,fm 
(open triangle), $2.5$\,fm (filled triangle), $2$\,fm (open diamond) and 
$1.8$\,fm (filled diamond).}
\label{fig:gA b}
\end{center}
\end{figure}

The results for the axial 
coupling constant $\gA$ are plotted against $\mpi^2$ in Fig.~\ref{fig:gA b}, 
with each line representing 
a different volume. 
We see that the results for 
different volumes converge at large $\mpi^2$, revealing very 
limited volume dependence in this region. This is what we expect 
to see because at large pion masses the pion field will die out quickly 
so that the effect of the boundary is minimal. 
At low $\mpi$ the axial coupling constant exhibits large finite volume 
effects.
Indeed, for the two lightest masses, $\mpi=140$\,MeV and $\mpi=200$\,MeV, the 
effect is very large, with $\gA$ decreasing by around $30\%$ and $50\%$ 
from $L=10$ to $2.5$\,fm in each case. 
It is also interesting to note the turn-around in the low $\mpi$ 
behaviour of $\gA$ between the $L=10$\,fm and $L=5$\,fm solutions. 

\section{Discussion}
Our plot of the hedgehog axial coupling constant reveals a large volume 
dependence in the low $\mpi$ region, with $\gA$ at the physical pion mass 
decreasing by over $50\%$ as the volume size decreases. 
This behaviour is similar to that observed in recent lattice QCD results.
From the plots of the hedgehog solution in \fig{hh solns} we see that the 
important region to consider is the distance $L/2-R$ (where we
once again use $L$ to denote the length of the side of the
lattice). This is the distance between 
the edge of the baryon and the volume boundary. Inside the baryon bag 
the fields do not vary much with volume size. 
We expect that this conclusion is far more general than the particular
model considered.

Clearly this has been a very simple study, involving a number of 
major approximations which must be remembered when considering our results. 
But by employing such a simple model we have been able to generate results 
over a wide range of volumes and pion masses with relative ease. 
Because of the exact nature of the solutions, at this point it would be 
very easy to examine the volume and mass dependence of other hedgehog 
properties. It would clearly be  
very interesting to employ a spontaneous symmetry breaking 
mechanism to include the pion and sigma masses. 

\subsection{Acknowledgments}
This work was supported by the Australian Research Council and by the
DOE under contract DE-AC05-84ER40150, under which SURA operates
Jefferson Lab.
\section{References}

\end{document}